\documentclass[preprint,aps,pra,showpacs,superscriptaddress,floatfix]{revtex4}
\usepackage{bm}
\usepackage{graphicx}  % Required for including images
\usepackage{braket}
\usepackage{booktabs} % Top and bottom rules for tables

\usepackage{amsmath}
\usepackage{exscale}
\usepackage{color}

\usepackage{ulem}

\usepackage[linktocpage=true,colorlinks=true,urlcolor=blue,plainpages=false,pdfpagelabels=false]{hyperref}
\begin{document}
\title{One-center calculations of the electron-positron pair creation in low-energy collisions of heavy bare nuclei} % Poster title

\author{R. V. Popov} 

\affiliation{ Department of Physics, St. Petersburg State University, 199034 St. Petersburg, Russia}
\affiliation{ NRC ``Kurchatov Institute'' -- ITEP, 117218 Moscow, Russia}

\author{A.~I.~Bondarev} % \writeto{bondarev@pcqnt1.phys.spbu.ru}

\affiliation{ Department of Physics, St. Petersburg State University, 199034 St. Petersburg, Russia}
\affiliation{ NRC ``Kurchatov Institute'' -- ITEP, 117218 Moscow, Russia}
\affiliation{ Center for Advanced Studies, Peter the Great St. Petersburg Polytechnic University, 195251 St. Petersburg, Russia}

\author{Y.~S.~Kozhedub} % \writeto{y.kozhedub@spbu.ru}

\affiliation{ Department  of Physics, St. Petersburg State University, 199034 St. Petersburg, Russia}

\author{I.~A.~Maltsev} % \writeto{ilia.alexm@gmail.com}

\affiliation{ Department of Physics, St. Petersburg State University, 199034 St. Petersburg, Russia}
\affiliation{ NRC ``Kurchatov Institute'' -- ITEP, 117218 Moscow, Russia}

\author{V.~M.~Shabaev} %\writeto{v.shabaev@spbu.ru}

\affiliation{ Department of Physics, St. Petersburg State University, 199034 St. Petersburg, Russia}

\author{I.~I.~Tupitsyn}

\affiliation{ Department of Physics, St. Petersburg State University, 199034 St. Petersburg, Russia}

\author{X.~Ma}

\affiliation{Institute of Modern Physics, Chinese Academy of Sciences, 730000  Lanzhou, China}

\author{G.~Plunien}
\affiliation{Institut f\"ur Theoretische Physik, Technische Universit\"at Dresden,
 D-01062 Dresden, Germany}

\author{Th.~St\"ohlker}
\affiliation{ GSI Helmholtzzentrum f\"ur Schwerionenforschung GmbH, D-64291 Darmstadt, Germany}
\affiliation{ Helmholtz-Institute Jena, D-07743 Jena, Germany}
\affiliation{ Institut f\"ur  Optik  und  Quantenelektronik, Friedrich-Schiller-Universit\"at, D-07743 Jena,  Germany}

\begin{abstract}
  The probabilities of bound-free electron-positron pair creation are calculated for head-on collisions
  of bare uranium nuclei beyond the monopole approximation. The calculations are
  based on the numerical solving of the time-dependent Dirac equation in the target reference
  frame with multipole expansion of the projectile potential.
  In addition, the energy dependence of 
  the pair-creation cross section is studied in the monopole approximation.
\end{abstract}
\maketitle

\section{Introduction}
\label{sec:1}

Spontaneous electron-positron pair creation in the presence of supercritical Coulomb field
is a fundamental effect of quantum electrodynamics, which was first predicted in Refs.~\cite{Gersh69,Piep69}.
Low-energy heavy-ion collisions can provide a field of the required strength and
therefore can serve as a tool for investigations of this phenomenon \cite{Grein85}.
The crucial condition imposed on the colliding nuclei is that their total charge $Z_{\rm tot}=Z_1+Z_2$
should exceed the critical value, $Z_{\rm cr}\approx 173$ (see Ref. \cite{Grein85} and references therein). 
However, the spontaneous contribution to the pair creation has to be distinguished from the dynamical one which
occurs due to the time dependence of the potential of the moving nuclei.
The pure spontaneous pair creation was investigated
in Refs.~\cite{Gershtein1973,Popov73,Peitz1973}. Analytical evaluation of the dynamical
contribution for  $\alpha Z_{1,2}  \ll 1$ ($\alpha$ is the fine structure constant) was carried out 
in Ref.~\cite{lee2016electron}. A rough estimate of this contribution for heavy ions
was considered in Refs. \cite{Khrip16,khriplovich2017positron}.
The nonperturbative consideration of pair creation with simultaneous inclusion
of both contributions requires solving the time-dependent Dirac equation (TDDE).
As applied to the pair-creation calculations, several techniques were utilized
\cite{Reinhardt81, Muller88,Ackad08,Malts15,bondarev2015positron} for solving the TDDE
in the monopole approximation. In this approximation, only the spherically symmetric part
of the two-center potential is taken into account. Usually the monopole approximation is
used in the center-of-mass~(CM) reference frame, because it provides better description of
the homonuclear quasi-molecule for small internuclear distances. However, for  large internuclear
distances the reference frame of one of the nuclei (target) is preferable.

The only way to test the validity of the monopole approximation is to go beyond it. In Ref.~\cite{Malts17}, calculations of pair creation with the full two-center potential were performed in the CM frame. Another way to go beyond the monopole approximation is to take into account the higher-order terms of the multipole expansion. This technique was used in Refs.~\cite{Rafelski76, Soff79, Marsman11, Roenko18} for solving the two-center stationary Dirac equation and in Refs.~\cite{mcconnell2012solution,mcconnell2013treatment, Bondarev15} for solving the TDDE as applied to calculations of ionization probabilities in heavy-ion collisions.   

In the present paper, we consider the collision process in the target reference frame. The target potential is fully accounted, whereas the projectile potential is expanded in the multipole series truncated at some order. The TDDE is solved using the finite basis set of hydrogenlike wave functions. The basis functions are constructed from B-splines using the dual-kinetic-balance approach (DKB)~\cite{Shab04}. The calculations of pair-creation probabilities are performed for the collision of bare uranium nuclei at energy near the Coulomb barrier. In order to reduce the computation time, we consider only the bound-free pair creation since the bound-state contribution is expected to be the dominant one~\cite{Muller88, Malts15}. The obtained results are compared with the corresponding values calculated with the full two-center potential~\cite{Malts17} and with the monopole-approximation
potential  in the CM frame~\cite{Muller88,Malts15}. We also  evaluate the cross section of pair creation for the collision of bare uranium nuclei at different energies
in the monopole approximation. These calculations are performed in the target as well as in the CM frames.

% Here some references that can be added are listed (.bib file is changed accordingly):
% 

Throughout the paper $\hbar = 1$ is assumed.

%----------------------------------------------------------------------------------------
%	GENERAL FORMALISM
%----------------------------------------------------------------------------------------

\section{Theory}
\label{sec:2}

In the present work, the nuclear collision process is treated semiclassically. Under this approximation, the colliding nuclei are regarded as the sources of an external time-dependent potential. Their motion is described classically with trajectories of the Rutherford type. The magnetic part of the potential is neglected due to the smallness of the relative collision velocity compared to the speed of light. 
The electron dynamics is described by the TDDE:
\begin{align}
     \label{eqn:psi_t}
     i\frac{\partial}{\partial t} \psi (\bm r, t) = H(t) \psi (\bm r, t),
\end{align}
where
\begin{align}
 H(t) = c(\bm{\alpha}\cdot\bm{p}) + \beta m_ec^2  + V_{\rm tot} \left(\bm r, t \right).
\end{align}
Here $\bm{\alpha}$, $\beta$ are the Dirac matrices, $c$ is the speed of light, $m_e$ denotes the electron mass, and $V_{\rm tot}$ is the total two-center potential of the colliding nuclei:
\begin{align}
 V_{\rm tot} \left(\bm r, t \right) = 
 V_{\rm T} \left(|\bm r - \bm R_{\rm T}(t)| \right) +
 V_{\rm P} (|\bm r - \bm R_{\rm P}(t)|),
\end{align}
where vectors $\bm R_{\rm T}$ and $\bm R_{\rm P}$ denote the positions of the target and projectile nuclei, respectively, and 
\begin{equation}
 V_{\rm T, P} (r) = 
 \int d \bm r^\prime \,
 \frac{\rho_{\rm \, T, P} (\bm r^\prime)}{|\bm r - \bm r^\prime|}
\end{equation}
are the corresponding nuclear potentials. For the nuclear charge distribution $\rho (\bm r)$ we utilize the model of the uniformly charged sphere.

Assuming the coordinate origin is on the internuclear axis,  the multipole expansion of the two-center potential can be written as
\begin{align}
 \label{eqn:mult_exp}
 V_{\rm tot} (\bm r, t) = 
 \sum\limits_{l = 0}^\infty 
 \left( 
 V_{\rm T}^l (r, R_{\rm T} (t)) +V_{\rm P}^l  (r, R_{\rm P} (t)) 
 \right)
 P_l (\cos \theta).
\end{align}
Here $P_l$ are the Legendre polynomials, $\theta$ is the angle between 
vectors $\bm r$ and $\bm R = \bm R_{\rm P} - \bm R_{\rm T}$, and 
\begin{align}
 V_{\rm T, P}^l (r, R_{\rm T, P}) = \frac{2l + 1}{2} \int\limits_0^\pi d(\cos \theta) V_{\rm T, P} (\bm r, \bm R_{\rm \, T, P}) P_l (\cos \theta). 
\end{align}
The expansion (\ref{eqn:mult_exp}) depends on the position of the coordinate origin. In the target reference frame, $R_{\rm T} = 0$ and $R_{\rm P} = R(t)$ is the internuclear distance. Then Eq.~(\ref{eqn:mult_exp}) gives
\begin{align}
 \label{eqn:mult_exp_target}
 V_{\rm tot} (\bm r, t) = 
 V_{\rm T} (r) + 
 \sum\limits_{l = 0}^\infty V_{\rm P}^l \left( r, R(t) \right) P_l (\cos \theta).
\end{align}
In the monopole approximation, only the term with $l = 0$ is taken into account and Eq.~(\ref{eqn:mult_exp_target})
is reduced to
\begin{align}
 V_{\rm tot} (\bm r, t) \simeq V_{\rm mon}^{\rm T} \left(r, R(t) \right) = 
 V_{\rm T} (r) +  V_{\rm P}^0 \left(r, R(t) \right).
\end{align}
In the CM frame, for two nuclei with equal masses the monopole approximation 
has the following form:
\begin{align}
 V_{\rm tot} (\bm r, t)  \simeq V_{\rm mon}^{\rm CM}  \left(r, R(t) \right) =
 V_{\rm T}^0 \left(r, R(t)/2 \right) +  V_{\rm P}^0 \left(r, R(t)/2 \right).
\end{align}
The monopole potentials in different reference frames have different asymptotics 
for $R \rightarrow \infty$: $V_{\rm mon}^{\rm T} (r, R) \rightarrow V_{\rm T}(r)$ and $V_{\rm mon}^{\rm CM} (r, R) \rightarrow 0$. Therefore only in the target frame the monopole Hamiltonian
has well-defined bound states for large internuclear distances.
However, the CM monopole potential is better in describing the two-center potential at small internuclear distances.
The monopole approximation allows us to reduce the three-dimensional TDDE to the one-dimensional equation that
drastically simplifies the numerical calculations. Adding the higher-order multipole terms, one should improve the approximation. 

We note that the target reference frame is non-inertial. However, we neglect the corresponding correction to the Hamiltonian assuming that its influence is small enough. 

 To describe the process of pair creation, the formalism of quantum electrodynamics with the unstable vacuum is employed \cite{EGitm91,Grein85}. Let us introduce two sets of solutions of the TDDE \eqref{eqn:psi_t}:  $\{\psi^{(+)}_n(\bm{r},t)\}$ are the in-solutions  and $\{\psi^{(-)}_n(\bm{r},t)\}$ are the out-solutions. The sets differ by the boundary conditions imposed on the wave function at the initial $t_{\rm in}$ and  final $t_{\rm out}$ time moments:
\begin{align}
  \psi^{(+)}_n(\bm{r},t_{\rm in}) = \phi_n(\bm{r}),\\    
  \psi^{(-)}_n(\bm{r},t_{\rm out}) = \phi_n(\bm{r}),
\end{align}
where $\phi_n$ are the solutions of the stationary Dirac equation
\begin{gather}
  H_0\phi_n = \varepsilon_n\phi_n,\\
  H_0 = c(\bm{\alpha}\cdot\bm{p}) + \beta m_ec^2  + U(\bm r).
\end{gather}
We assume that $H(t_{\rm in}) = H(t_{\rm out}) = H_0$ since $R(t_{\rm in}) = R(t_{\rm out})$ in our calculations.
The sets  $\{\psi^{(+)}_n\}$ of in-solutions and $\{\psi^{(-)}_n\}$ of out-solutions describe physical particles at
times $t_{\rm in}$ and $t_{\rm out}$, correspondingly. We have chosen~$U(\bm r) = V_{\rm T}(r)$ for
the calculations in the target frame and~$U(\bm r) = V_{\rm mon}^{\rm CM} (r, t_{\rm in})$ for the calculations in
the CM frame within the monopole approximation. 

The mean number $n_m$ of electrons created from the vacuum in the state $m$  is given by
\begin{align}
  \label{eqn:sprob}
  n_m = \sum_{n < { F}}\left|a_{mn}\right|^2,
\end{align}
where ${ F}$ is the Fermi level ($\varepsilon_{\rm F} = - m_e c^2$) and the one-electron transition amplitudes $a_{mn}$ are defined as
\begin{align}
  \label{eqn:1pamp0}
  a_{mn} = \int {\rm d}\bm{r} \; \psi^{(-)}_m{}^{\dag}(\bm{r},t)\psi^{(+)}_n(\bm{r},t).
\end{align}
The amplitudes $a_{mn}$ are time-independent~\cite{Malts15}, hence one can consider them at the time moment~$t_{\rm in}$:
\begin{align}
  a_{mn} = \int {\rm d}\bm{r} \; \psi^{(-)}_m{}^{\dag}(\bm{r},t_{\rm in}) \phi_n(\bm r).
\end{align}
The wave functions $\psi^{(-)}_m$ at the time moment $t_{\rm in}$ are found using the numerical solution of the TDDE.
The initial states $\phi_n$, including the bound ones and the pseudostates from both (negative- and positive-energy) continuum spectra,
are obtained by diagonalization of the $H_0$ matrix in a finite basis set.
The basis functions are generated from the B-splines according to the DKB technique~\cite{Shab04}.
The time-dependent wave functions are decomposed over the obtained $\phi_n$ states:
\begin{align}
\label{eqn:PhiDecomp}
  \psi_i(\bm{r},t) = \sum_{k = 1}^N c_{ki}(t)\phi_k(\bm{r})e^{-i\varepsilon_kt},
\end{align}
where $N$ is the number of the states, $\varepsilon_k$ are the eigenvalues of the $H_0$ matrix, and $c_{ki}$ are the expansion coefficients. The representation \eqref{eqn:PhiDecomp} leads to the system of differential equations on the expansion  coefficients:
\begin{align}
  \label{eqn:SDEq}
    i \frac{\partial}{\partial t} c_{ji}(t) = \sum_k V_{jk}(t)c_{ki}(t), \text{\ \ \ subject to } c_{ji}(t_{\rm in}) = \delta_{ji},
\end{align}
where 
\begin{align}
  \label{eqn:Vij}
  V_{jk}(t) = \langle \phi_j |(V_{\rm tot}(t) - U)| \phi_k \rangle  e^{-i(\varepsilon_k - \varepsilon_j)t}. 
\end{align}
In the target frame, $V_{\rm tot}(t) - U = V_{\rm P} (t)$. In order to calculate the matrix elements $V_{jk}$, the target potential $V_{\rm P} (t)$ is expanded in the multipole series according to Eq.~\eqref{eqn:mult_exp_target} and the expansion is truncated at some order.

The system of equations~\eqref{eqn:SDEq} is solved employing the Crank-Nicolson scheme~\cite{Crank47}:
%\begin{align}
%\vec{c}_i(t + \Delta t) \approx M(t + \Delta t; t)\vec{c}_{i}(t),
%\end{align}
\begin{align}
  {\vec c}_i(t + \Delta t) \approx M(t + \Delta t; t){\vec c}_{i}(t),
\end{align}
where $\Delta t$ is a sufficiently small time step, 
$\vec c_i = \{c_{1i}, \dots, c_{Ni} \}$,
and the matrix $M$ is defined as
\begin{align}
  M(t + \Delta t;\; t) = \left[I + i \frac{\Delta t}{2}V(t + \frac{\Delta t}{2})\right]^{-1}
  \left[I - i \frac{\Delta t}{2}(t + V\frac{\Delta t}{2})\right].
\end{align}
Using the described technique one can propagate all the bound states back in time from $t_{\rm out}$ to $t_{\rm in}$ and calculate the total bound-free pair-creation probability,
\begin{align}
 P_b = \sum_{|\varepsilon_k| < m_e c^2} n_k,
\end{align}
as well as the final electron population of each bound state.
The pair-creation cross section can be found by integration over the impact parameter $b$,
\begin{align}
  \sigma_b = 2\pi \int\limits_{0}^\infty db \, b \, P_b.
\end{align}

% --------------------------------------------------------------------------------
% RESULTS
% --------------------------------------------------------------------------------

\section{Results}
\label{sec:4}

\begin{figure}
  \centering
  \caption{Electron-positron pair-creation cross section with electron captured into a bound state, calculated
    within the monopole approximation.
    %using a basis set which contains functions with zero orbital angular momentum only.
    Red triangles (green circles) indicate the results obtained in the target (center-of-mass) reference frame.}
  \label{fig:1}
  \includegraphics[width=15.0cm]{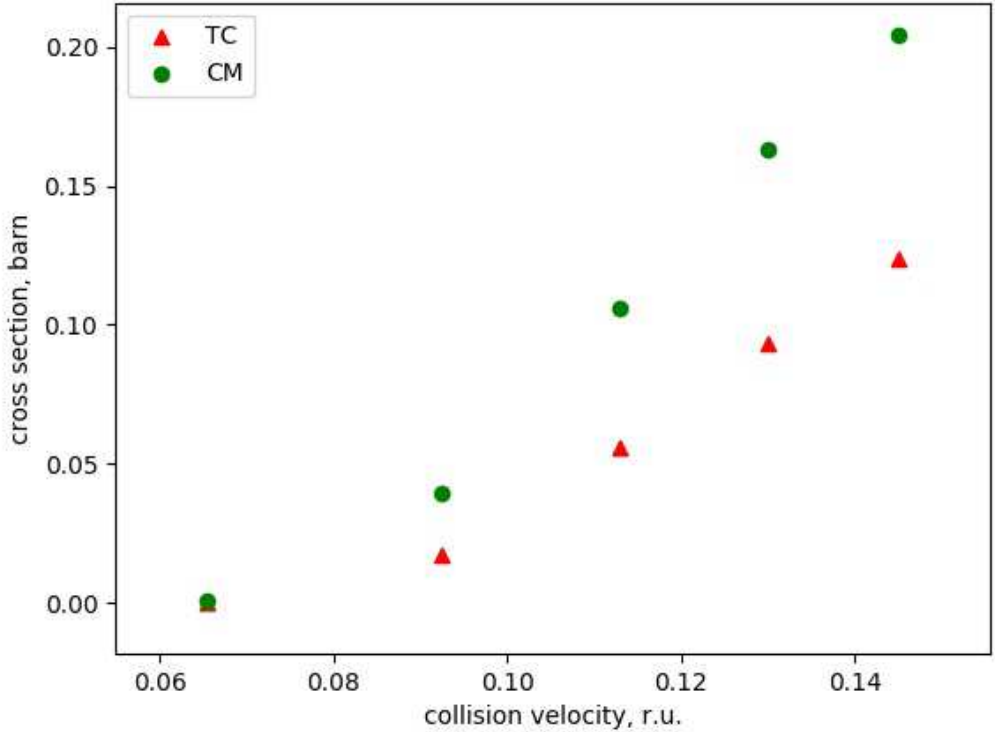}
\end{figure}

Employing the method described above, we performed the calculations
of electron-positron pair-creation probabilities in collisions of bare uranium nuclei. The pair-creation cross sections were obtained in the monopole approximation with the basis set which includes functions with zero orbital angular momentum only.
The calculations were carried out in the target and CM reference frames for a wide range of the collision energies.
The results are depicted in Fig.~\ref{fig:1} as functions of the asymptotic collision velocity.
The obtained CM values are systematically larger than the target ones. This can be explained by the fact
that at small internuclear distances the CM monopole potential
is stronger than the target one. The cross section calculated for the collision velocity near $0.1$ relativistic unit (r.u.) 
is about two orders of magnitude smaller than the value obtained by a rough
estimate in Ref.~\cite{khriplovich2017positron}.

Next, we performed the calculations beyond the monopole approximation for the head-on collisions.
For the direct comparison of our results with the data of Refs.~\cite{Malts17,Malts15},
the collision energy $E=6.218$ MeV/u was used. Table~\ref{tab:1} represents the pair-creation probabilities
obtained with truncation of the time-dependent wave function decomposition \eqref{eqn:PhiDecomp}
at different orbital momenta.
The results of the full two-center calculation~\cite{Malts17} and the value obtained within the monopole approximation
in the CM frame~\cite{Malts15} are also presented.
Despite our approach is too rough to account properly for the  process of pair creation with
electron capture by the projectile, the obtained values are very close to ones of Ref.~\cite{Malts17},
where this process is embedded in the calculation technique. 
A possible explanation for this could be as follows:  the electron-positron pairs
are mainly created at small internuclear distances where the higher-order
multipole terms of the projectile potential do not play a significant role.  
\begin{table}[h]
  \centering
  \caption{Probability of pair creation with electron captured into the ground state ($P_{\rm g}$) and into
    any bound state ($P_{\rm b}$) in U$^{92+}$-U$^{92+}$ head-on collisions at energy $E=6.218$ MeV/u. 
    Here $l_{\rm max}$ is the maximal orbital momentum of the wave functions included in the basis set.
    For comparison, the results obtained in Ref.~\cite{Malts15} with the CM monopole potential and the values of Ref.~\cite{Malts17} calculated with the full two-center potential are also presented.}
  \label{tab:1}
  \begin{tabular}{c*{2}{|c}}
    \hline
    \hline
 
    $l_{\rm max}$ & $P_{\rm g}$  &  $P_{\rm b}$ \\
    \hline
    $0$          & $5.73\cdot 10^{-3}$  & $5.91\cdot 10^{-3}$ \\ 
    $1$          & $9.23\cdot 10^{-3}$  & $1.05\cdot 10^{-2}$ \\
    $2$          & $1.05\cdot 10^{-2}$  & $1.24\cdot 10^{-2}$ \\
    $3$          & $1.10\cdot 10^{-2}$  & $1.30\cdot 10^{-2}$ \\
    $4$          & $1.11\cdot 10^{-2}$  & $1.31\cdot 10^{-2}$ \\
    $5$          & $1.09\cdot 10^{-2}$  & $1.29\cdot 10^{-2}$ \\
    \hline
    CM~monopole~\cite{Malts15}& &$1.25 \cdot 10^{-2}$\\
    \hline
    two-center~\cite{Malts17} & $1.11\cdot 10^{-2}$  & $1.32\cdot 10^{-2}$  \\
    \hline
    \hline
  \end{tabular}
\end{table}

It should be noted that in our calculations we restricted the basis set to states
with the total angular momentum projection on the internuclear axis  $\mu$ equal to $\pm 1/2$ only.
In the head-on collision, the time-dependent potential does not mix states with different $\mu$ values.
Thus, the contributions of these states can be calculated independently. Moreover, the contributions of the states
with opposite angular momentum projections are equal to each other.
Hence, it is sufficient to carry out the evaluation for the basis set with a certain sign of
the angular momentum projection and double the obtained value. As in Ref.~\cite{Malts17}, it is found that the contribution of
states with the angular momentum projection larger than $1/2$ is negligible.  

\section{Conclusion}
\label{sec:Concl}

In this paper we have evaluated the electron-positron pair-creation probabilities for the head-on collision of bare uranium nuclei at the energy $E=6.218$ MeV/u beyond the monopole approximation. 
The calculations were performed using one-center basis set expansion in the target reference frame.
The target potential was fully taken into account, while the potential of the projectile was
approximated by few lowest-order terms of the multipole expansion
with respect to the target nucleus. 
The results of the calculations are in reasonable agreement with
the data obtained within the framework of the full two-center potential approach~\cite{Malts17}.
Further improvement of the accuracy by adding the higher-order terms is limited by
computational resources.

We  have also calculated the pair-creation cross section in the monopole approximation for collision
energies in the range of 2--10 MeV/u. The calculations were carried out in
the target as well as in the CM frames. The obtained values have the same order of magnitude
in both frames.

\section{Acknowledgment}
\label{sec:Ack}

This work was supported by RFBR-NSFC (Grants No. 17-52-53136 and No. 11611530684),
by SPbSU-DFG (Grants No. 11.65.41.2017  and No. STO 346/5-1), by SPbSU (Grant No. 11.40.538.2017), 
and by the Ministry of Education and Science of the Russian Federation (Grant No. 3.1463.2017/4.6).
R.V.P., A.I.B., and I.A.M. also acknowledge the support of the FAIR-Russia Research Center. 
The work of V.M.S. was supported by the CAS President International Fellowship Initiative (PIFI).


\footnotesize

\begin{thebibliography}{99}
%########################

\bibitem{Gersh69}
S.S.~Gershtein and Y.B.~Zeldovich, Zh. Eksp. Teor. Fiz. {\bf 57}, 654 (1969) [Sov. Phys. JETP {\bf 30}, 358 (1970)].

%%%%%%%%%%%%%%%%%%%%%%%%%

\bibitem{Piep69}
W.~Pieper and W.~Greiner, Z. Physik {\bf 218}, 327 (1969).

%%%%%%%%%%%%%%%%%%%%%%%%%

\bibitem{Grein85}
W.~Greiner, B.~M\"uller, and J.~Rafelski, {\it Quantum Electrodynamics of Strong Fields} (Springer-Verlag, Berlin, 1985).

%%%%%%%%%%%%%%%%%%%%%%%%%

\bibitem{Gershtein1973}
S.S.~Gershtein and V.S.~Popov, \href{https://doi.org/10.1007/BF02827078}{Lett. Nuovo Cimento {\bf 6}, 593 (1973).}

%%%%%%%%%%%%%%%%%%%%%%%%%

\bibitem{Popov73}
V.S.~Popov, Zh. Eksp. Teor. Fiz. {\bf 65}, 35 (1973) [Sov. Phys. JETP {\bf 38}, 18 (1974)].

%%%%%%%%%%%%%%%%%%%%%%%%%

\bibitem{Peitz1973}
H.~Peitz, and B.~M{\"u}ller, and J.~Rafelski, and W.~Greiner, \href{https://doi.org/10.1007/BF02727627}{Lett. Nuovo Cimento {\bf 8}, 37 (1997).}

%%%%%%%%%%%%%%%%%%%%%%%%%

\bibitem{lee2016electron}
R.N.~Lee and A.I.~Milstein, \href{https://doi.org/10.1016/j.physletb.2016.08.058}{Phys. Lett. B {\bf 761}, 340 (2016).}

%%%%%%%%%%%%%%%%%%%%%%%%%
\bibitem{Khrip16}
I.B.~Khriplovich, \href{http://www.worldscientific.com/doi/abs/10.1142/S0217751X16450354}{Int. J. Mod. Phys. A {\bf 31}, 1645035 (2016).}
%%%%%%%%%%%%%%%%%%%%%%%%%%
\bibitem{khriplovich2017positron}
I.B.~Khriplovich, \href{https://doi.org/10.1140/epjp/i2017-11329-8}{Eur. Phys. J. Plus {\bf 132}, 61 (2017).}

%%%%%%%%%%%%%%%%%%%%%%%%%

\bibitem{Reinhardt81}
J.~Reinhardt, B.~M\"uller, and W.~Greiner, \href{https://link.aps.org/doi/10.1103/PhysRevA.24.103}{Phys. Rev. A {\bf 24}, 103 (1981).}

%%%%%%%%%%%%%%%%%%%%%%%%%

\bibitem{Muller88}
U.~M\"uller, T.~de~Reus, J.~Reinhardt, B.~M\"uller, W.~Greiner, and G.~Soff, \href{https://link.aps.org/doi/10.1103/PhysRevA.37.1449}{Phys. Rev. A {\bf 37}, 1449 (1988).}

%%%%%%%%%%%%%%%%%%%%%%%%%

\bibitem{Ackad08}
E.~Ackad and M.~Horbatsch, \href{https://link.aps.org/doi/10.1103/PhysRevA.78.062711}{Phys. Rev. A {\bf 78}, 062711 (2008).}

%%%%%%%%%%%%%%%%%%%%%%%%%

\bibitem{Malts15}
I.A.~Maltsev, V.M.~Shabaev, I.I.~Tupitsyn, A.I.~Bondarev, Y.S.~Kozhedub, G.~Plunien, and T.~St\"ohlker, \href{https://link.aps.org/doi/10.1103/PhysRevA.91.032708}{Phys. Rev. A {\bf 91}, 032708 (2015).}

%%%%%%%%%%%%%%%%%%%%%%%%%

\bibitem{bondarev2015positron}
A.I.~Bondarev, I.I.~Tupitsyn, I.A.~Maltsev, Y.S.~Kozhedub, and G.~Plunien, \href{https://doi.org/10.1140/epjd/e2015-50783-6}{Eur. Phys. J. D {\bf 69}, 110 (2015).}

%%%%%%%%%%%%%%%%%%%%%%%%%

\bibitem{Malts17}
I.A.~Maltsev, V.M.~Shabaev, I.I.~Tupitsyn, Y.S.~Kozhedub, G.~Plunien, and T. St\"ohlker, \href{http://www.sciencedirect.com/science/article/pii/S0168583X17305797}{Nucl. Instrum. Methods Phys. Res. B {\bf 408}, 97 (2017).}

%%%%%%%%%%%%%%%%%%%%%%%%%

\bibitem{Rafelski76}
J.~Rafelski and B.~M\"uller, \href{https://link.aps.org/doi/10.1103/PhysRevLett.36.517}{Phys. Rev. Lett. {\bf 36}, 517 (1976).}

%%%%%%%%%%%%%%%%%%%%%%%%%

\bibitem{Soff79}
G.~Soff, W.~Greiner, W.~Betz, and B. M\"uller, \href{https://link.aps.org/doi/10.1103/PhysRevA.20.169}{Phys. Rev. A {\bf 20}, 169 (1979).}

%%%%%%%%%%%%%%%%%%%%%%%%%

\bibitem{Marsman11}
A.~Marsman and M.~Horbatsch, \href{https://link.aps.org/doi/10.1103/PhysRevA.84.032517}{Phys. Rev. A {\bf 84}, 032517 (2011).}

%%%%%%%%%%%%%%%%%%%%%%%%%

\bibitem{Roenko18}
A.~Roenko and K.~Sveshnikov, \href{https://link.aps.org/doi/10.1103/PhysRevA.97.012113}{Phys. Rev. A {\bf 97}, 012113 (2018).}

%%%%%%%%%%%%%%%%%%%%%%%%%

\bibitem{mcconnell2012solution}
S.R.~McConnell, A.N.~Artemyev, M.~Mai, and A.~Surzhykov, \href{https://doi.org/10.1103/PhysRevA.86.052705}{Phys. Rev. A {\bf 86}, 052705 (2012).}

%%%%%%%%%%%%%%%%%%%%%%%%%

\bibitem{mcconnell2013treatment}
S.R.~McConnell, A.N.~Artemyev, and A.~Surzhykov, \href{http://iopscience.iop.org/article/10.1088/0031-8949/2013/T156/014056}{Phys. Scr. {\bf T156}, 014055 (2013).}

%%%%%%%%%%%%%%%%%%%%%%%%%

\bibitem{Bondarev15}
A.I.~Bondarev, I.V.~Ivanova, I.I.~Tupitsyn, Y.S.~Kozhedub, and G.~Plunien, \href{http://stacks.iop.org/1742-6596/635/i=2/a=022094}{J. Phys.: Conf. Ser. {\bf 635}, 022094 (2015).}

%%%%%%%%%%%%%%%%%%%%%%%%%

\bibitem{Shab04}
V.M.~Shabaev, I.I.~Tupitsyn, V.A.~Yerokhin, G.~Plunien, and G.~Soff, \href{https://link.aps.org/doi/10.1103/PhysRevLett.93.130405}{Phys. Rev. Lett. {\bf 93}, 130405 (2004).}

%%%%%%%%%%%%%%%%%%%%%%%%%

\bibitem{EGitm91}
 E.S.~Fradkin, D.M.~Gitman, and S.M.~Shvartsman, {\it Quantum Electrodynamics with Unstable Vacuum} (Springer-Verlag, Berlin, 1991).

%%%%%%%%%%%%%%%%%%%%%%%%%

\bibitem{Crank47}
J.~Crank and P.~Nicolson, \href{https://doi.org/10.1017/S0305004100023197}{Proc. Cambridge Philos. Soc. {\bf 43}, 50 (1947).} 

%%%%%%%%%%%%%%%%%%%%%%%%%


%########################
\end{thebibliography}
\end{document}